# A Robust and Secure Aggregation Protocol for Wireless Sensor Networks

Jaydip Sen
Innovation Lab, Tata Consultancy Services Ltd.
Bengal Intelligent Park, Salt Lake Electronic Complex
Kolkata- 700091, INDIA
email: Jaydip.Sen@tcs.com

*Abstract*—The purpose of a wireless sensor network (WSN) is to provide the users with access to the information of interest from data gathered by spatially distributed sensors. Generally the users require only certain aggregate functions of this distributed data. Computation of this aggregate data under the end-to-end information flow paradigm by communicating all the relevant data to a central collector node is a highly inefficient solution for this purpose. An alternative proposition is to perform in-network computation. This, however, raises questions such as: what is the optimal way to compute an aggregate function from a set of statistically correlated values stored in different nodes; what is the security of such aggregation as the results sent by a compromised or faulty node in the network can adversely affect the accuracy of the computed result. In this paper, we have presented an energy-efficient aggregation algorithm for WSNs that is secure and robust against malicious insider attack by any compromised or faulty node in the network. In contrast to the traditional snapshot aggregation approach in WSNs, a node in the proposed algorithm instead of unicasting its sensed information to its parent node, broadcasts its estimate to all its neighbors. This makes the system more fault-tolerant and increase the information availability in the network. The simulations conducted on the proposed algorithm have produced results that demonstrate its effectiveness.

## I. INTRODUCTION

The purpose of traditional data networks such as the Internet is to enable end-to-end information transfer. Information streams in such networks are carried across point-to-point links, with intermediate nodes simply forwarding data packets without modifying their payloads. In contrast, the purpose of a wireless sensor network (WSN) is to provide the users with access to the information of interest from the data gathered by spatially distributed sensors. In most applications, users require only certain aggregate functions of this distributed data. Examples include the average temperature in a network of temperature sensors, a particular trigger in the case of an alarm network, or the location of an event. Such aggregate functions could be computed under the end-to-end information flow paradigm by communicating all relevant data to a central collector node. This, however, is a highly inefficient solution for WSNs which have severe constraints in energy, memory and bandwidth, and where tight latency constraints are to be met. An alternative solution is to perform in-network computations [1]. However, in this case, the question that arises is how to perform distributed computations over a network of nodes connected by wireless links in an efficient manner. What is the optimal way to compute, for example, the average, min, or max of a set of statistically correlated values stored in different nodes? How would such computations be performed in the presence of unreliability such as noise, packet drops, and node failures? Such questions combine the complexities of multi-terminal information theory, distributed source coding, communication complexity, and distributed computation. This makes development of an efficient in-network computing framework for WSNs very challenging.

In this paper, we have considered a WSN as a collective entity that performs a sensing task and have proposed a distributed estimation algorithm that can be applied to a large class of aggregation problems. Apart from making a trade-off between the level of accuracy in aggregation and the energy expended in computation of the aggregate function, we have brought in another very important and relevant factor in WSN- security. Unfortunately, even though security has been identified as a major challenge for sensor networks [2], current proposals for data aggregation protocols have not been designed with security in mind, and consequently they are all vulnerable to easy attacks. Even when a single sensor node is captured, compromised or spoofed, an attacker can often manipulate the value of an aggregate function without any bound, gaining complete control over the computed aggregate. In fact, any protocol that computes the average, sum, minimum, or maximum function is insecure against malicious data, no matter how these functions are computed. Keeping in mind these threats, we have developed an energy-efficient aggregation algorithm that is secure and robust against malicious attacks in WSNs. The main threat that we have considered while designing the proposed scheme is the injection of malicious data in the network by an adversary who has compromised a sensor's sensed value by subjecting it to unusual temperature, lighting, or other spoofed environmental conditions.

In the proposed scheme, each node in a WSN has complete information about the parameter being sensed. This is in contrast to the snapshot aggregation, where the sensed parameters are aggregated at the intermediate nodes till the final aggregated result reaches the root. Each node, in the proposed algorithm, instead of unicasting its sensed information to its parent, broadcasts its estimate to all its neighbors. This makes the protocol more fault-tolerant and increases the information availability in the network. The proposed protocol is similar to the one suggested in [3]. However, it is more secure and reliable even in presence of compromised and faulty nodes in a WSN.

The rest of the paper is organized as follows. Section II presents some related work in the area of aggregation algorithms for WSNs. Section III discusses in details the proposed distributed estimation algorithm. Section VI presents the simulation results. Section V concludes the paper and also highlights some future scope of work.

## II. RELATED WORK

Extensive work has been done on aggregation applications in WSNs. However, security and energy- two major aspects for design of an efficient and robust aggregation algorithm have not attracted adequate attention. In [4] and [5] the authors have proposed a framework for flexible aggregation in WSNs. However, these propositions are based on snapshot aggregation and have not addressed the issues related to energy efficiency and security.

The authors in [6] have proposed a snapshot aggregation algorithm for nodes in an ad hoc network where each node has TinyOS as the operating system. The main contribution of the paper is the development of an interface for executing the snapshot aggregation. A query processing system is also presented for extracting information from the nodes in the network. A significant performance improvement is claimed as compared to traditional, centralized approaches of aggregation. However, the authors have considered energy efficiency and security aspects of computation of the aggregate function. The conventional aggregates like minimum, maximum, average, count, sum etc are all vulnerable to insider attacks by compromised or faulty nodes and thus the query systems based on those aggregates are not reliable. Propositions based on programmable sensor networks such as [7] also consider aggregation based on snapshot and therefore these schemes are inherently inefficient and not fault-tolerant.

In [8], the authors have focussed their attention into the problem of providing a residual energy map of a WSN. They have proposed a scheme for computing the equi-potential curves of residual energy with certain acceptable margin of error. A simple but efficient aggregation function is proposed where the location approximation of the nodes are not computed. A more advanced aggregate function can be developed for this purpose that will encompass an accurate convex curve. For periodic update of the residual energy map, the authors have proposed a naïve scheme of incremental updates. Thus if a node changes its value beyond the tolerance limit its value is transmitted and aggregated again by some nodes before the final change reaches the user. No mechanism exists for prediction of changes or for estimation of correlation between sensed values for the purpose of setting the tolerance threshold.

In [9], the authors have proposed a scheme for the purpose of monitoring the sensed values of each individual sensor node. There is no aggregation algorithm in the scheme; however, the spatial-temporal correlation between the sensed data can be extrapolated to fit an aggregation function. The authors have also attempted to modify the techniques of MPEG-2 for sensor network monitoring to optimize communication overhead and energy. A central node is computes predictions and transmits them to all the nodes. The nodes send their update only if their sensed data deviate significantly from the predictions. A distributed computing framework is developed by establishing a hierarchical dependency among the nodes.

An energy efficient aggregation algorithm is proposed by the authors in [3]. The scheme is a distributed estimation algorithm where each node in the network senses the parameter and there is no hierarchical dependency among the nodes. Nodes in a neighborhood periodically broadcast their information based on a threshold value. However, the scheme does not consider the security aspect of the aggregation algorithm. In this paper, we have extended the distributed estimation algorithm to make it secure and robust in presence of compromised and faulty nodes in a WSN.

## III. DIATRIBUTED AGGREGATION ALGORITHM

In this section, we propose the modified distributed estimation algorithm that is secure and resistant to insider attack by compromised and faulty nodes. There are essentially two categories of aggregation functions [3]: (i) aggregation functions that are dependent on the values of a few nodes (e.g., the *max* result is based on one node), and (ii) aggregation functions whose values are determined by all the nodes (e.g., the average function). However, computation of both these types of functions are adversely affected by wrong sensed result sent by even a very few number of compromised nodes. In this paper, we consider only the first case, i.e., aggregation function that find or approximate some kind of boundaries (e.g., maxima, minima), and hence the aggregation result is determined by the values of few nodes. However, the proposed algorithm does not assume any knowledge about the underlying physical process.

### A. Distributed Cooperative Approach

In the proposed distributed estimation algorithm, a sensor node instead of transmitting a partially aggregated result, maintains and if required, transmits an estimation of the global aggregated result. The global aggregated description in general will be a vector since it represents multi-dimensional parameters sensed by different nodes. A global estimate will thus be a probability density function of the vector that is being estimated. However, in most of the practical situations, due to lack of sufficient information, complex computational requirement or unavailability of sophisticated estimation tools, an estimate is represented as:

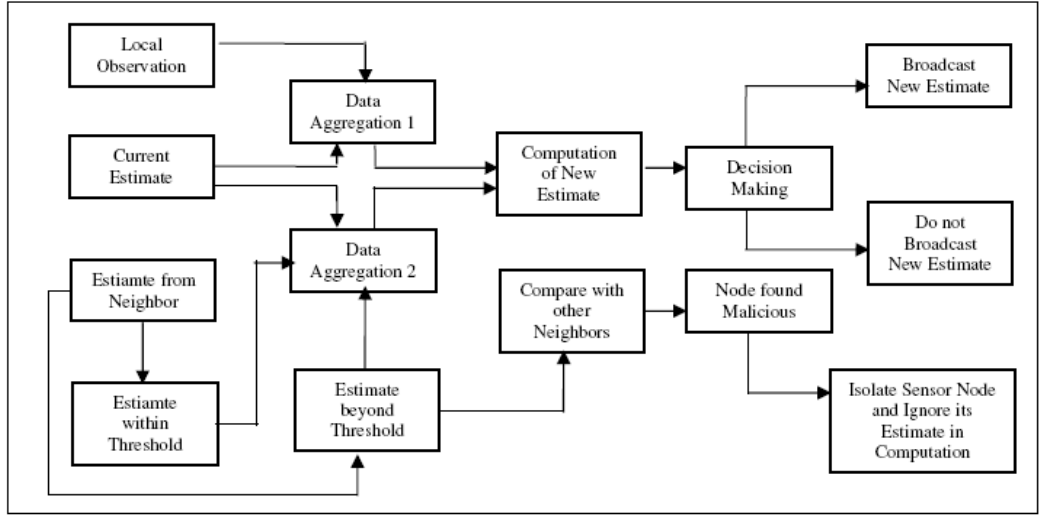

Figure 1. The schematic flow diagram of the proposed data aggregation algorithm for WSNs

(*estimated value*, *confidence indication*), which in computational terms can be represented as: (*average of estimated vector*, *covariance matrix of estimated vector*). For the sake of manipulability with tools of estimation theory, we have chosen to represent estimates in the form of ($A$, $P_{AA}$) with $A$ being the mean of the aggregated vector and $P_{AA}$ being the covariance matrix of vector $A$. For the *max* aggregation function, vector $A$ becomes a scalar denoting the mean of the estimated max, and $P_{AA}$ becomes simply the variance of $A$.

In the snapshot aggregation, a node does not have any control on the rate at which it send information to its parents; it has to always follow the rate specified the user application. Moreover, every node has little information about the global parameter, as it has no idea about what is happening beyond its parent. In proposed approach, a node accepts estimations from all of its neighbors, and gradually gains in knowledge about the global information. It helps a node to understand whether its own information is useful to its neighbors. If a node realizes that its estimate could be useful to its neighbors, it transmits the new estimate. Unlike snapshot aggregation where the node transmits its estimate to its parent, in the proposed scheme, the node broadcasts its estimate to all its neighbors. Moreover, there is no need to establish and maintain a hierarchical relationship among the nodes in the network. This makes the algorithm particularly suitable for multiple user, mobile users, faulty nodes and transient network partition situations.

### B. The Proposed Algorithm

The algorithm has the following steps:

1. Every node has an estimate of the global aggregated value (global estimate) in the form of (mean, covariance matrix). When a node makes a new local measurement, it makes an aggregation of the local observation with its current estimate. This is depicted in the block *Data Aggregation 1* in Fig.1. The node computes the new global estimate and decides whether it should broadcast the new estimate to its neighbors. The decision is based on a threshold value as explained in Section III E.

2. When a node receives a global estimate from a neighbor, it first checks whether the newly received estimate differs from its current estimate by more than a pre-defined threshold.

(a) If the difference does not exceed the threshold, the node makes an aggregation of the global estimates (its current value and the received value) and computes a new global estimate. This is depicted in the block *Data Aggregation 2* in Fig. 1. The node then decides whether it should broadcast the new estimate.

(b) If the difference exceeds the threshold, the node performs the same function as in step (a). Additionally, it requests its other neighbors to send their values of the global estimate.

(c) If the estimates sent by the majority of the neighbors differ from the estimate sent by the first neighbor by a threshold value, then the node is assumed to be compromised. Otherwise, it is assumed to be normal.

3. If a node is identified to be compromised, the global estimate previously sent by it is ignored in the computation of the new global estimate and the node is isolated from the network by a broadcast message in its neighborhood.

### C. Aggregation of Two Global Estimates

In Fig.1, the block *Data Aggregation 1* corresponds to this activity. For combining two global estimates to produce a single estimate, *covariance intersection* (CI) algorithm is used. CI algorithm is particularly suitable for this purpose, since it has the capability of aggregating two estimates without requiring any prior knowledge about their degree of correlation [10]. This is more pertinent to WSNs, as we cannot guarantee statistical independence of observed data in such networks.

Given two estimates ($A$, $P_{AA}$) and ($B$, $P_{BB}$), the combined estimate ($C$, $P_{CC}$) by CI is given by the following equations:

$$P_{CC} = (\omega * P_{AA}^{-1} + (1-\omega)P_{BB}^{-1})^{-1} \quad (1)$$

$$C = P_{CC}(\omega * P_{AA}^{-1} * A + (1-\omega)P_{BB}^{-1} * B) \quad (2)$$

Here, $P_{AA}$, $P_{BB}$, and $P_{CC}$ represent the covariance matrices associated with the estimates $A$, $B$, and $C$ respectively. The main computational problem with CI is the computation of $\omega$. The value of $\omega$ lies between 0 and 1. The optimum value of $\omega$ is arrived at when the trace of the determinant of $P_{CC}$ is minimized.

For *max* aggregation function, covariance matrices are simple scalars. It can be observed from Eqs. (1) and (2) that in such a case $\omega$ can be either 1 or 0. Subsequently, $P_{CC}$ is equal to the minimum of $P_{AA}$ and $P_{BB}$, and $C$ is equal to either $A$ or $B$ depending on the value of $P_{CC}$. Even when the estimates are reasonably small-sized vectors, there are efficient algorithms to determine $\omega$.

### D. Aggregation of Local and Global Observations

This module corresponds to the block *Data Aggregation 2* in Fig.1. Aggregation of a local observation with a global estimate involves a statistical computation with two probability distributions.

*Case 1*: Mean of the local observation is greater than the mean of the current global estimate: In case of *max* aggregation function, if the mean of the local observation is greater than the mean of the current global estimate, the local observation is taken as the new estimate. The distribution of the new estimate is arrived at by multiplying the distribution of the current global estimate by a positive fraction ($w_1$) and summing it with the distribution of the local observation. The fractional value determines the relative weight assigned to the value of the global estimate. The weight assigned to the local observation being unity.

*Case 2*: Mean of the local observation is smaller than the mean of the current global estimate: If a node observes that the mean of the local observation is smaller than its current estimate, it combines the two distributions in the same way as in *Case 1* above, but this time a higher weight ($w_2$) is assigned to the distribution having the higher mean (i.e. the current estimate). However, as observed in [3], this case should be handled more carefully if there is a sharp fall in the value of the global maximum. We follow the same approach as proposed in [3]. If the previous local measurement does not differ from the global estimate beyond a threshold value, a larger weight is assigned to the local measurement as in Case 1. In this case, it is believed that the specific local measurement is still the global aggregated value.

For computation of the weights $w_1$ and $w_2$ in *Case 1* and *Case 2* respectively, we follow the same approach as suggested in [3]. Since all the local measurements and the global estimates are assumed to follow Gaussian distribution, almost all the observations are bounded within the interval [$\mu \pm 3*\sigma$]. When the mean of the local measurement is larger than the mean of the global estimate, the computation of the weight ($w_1$) is done as follows. Let us suppose that $l(x)$ and $g(x)$ are the probability distributions for the local measurement and the global estimate respectively. If $l(x)$ and $g(x)$ can take non-zero values in the intervals [$x_1$, $x_2$] and [$y_1$, $y_2$] respectively, then the weight $w_1(x)$ will be assigned a value of 0 for all $x \leq \mu_1 - 3*\sigma$ and $w_1(x)$ will be assigned a value of 1 for all $x > \mu_1 - 3*\sigma$. Here, $x_1$ is equal to $\mu_1 - 3*\sigma_1$, where $\mu_1$ and $\sigma_1$ are the mean and the standard deviation of $l(x)$ respectively.

When the mean of the local measurement is smaller than the mean of the global estimate, the computation of the weight $w_2$ is carried out as follows. The value of $w_2(x)$ is assigned to be 0 for all $x \leq max\{\mu_1 - 3*\sigma_1, \mu_2 - 3*\sigma_2\}$. $w_2(x)$ is assigned a value of 1 for all $x > max\{\mu_1 - 3*\sigma_1, \mu_2 - 3*\sigma_2\}$. Here, $y_1$ is equal to $\mu_2 - 3*\sigma_2$, where $\mu_2$ and $\sigma_2$ represent the mean and the standard deviation of $g(x)$ respectively.

In all these computations, it assumed that resultant distribution after combination of two bounded Gaussian distributions is also a Gaussian distribution. This is done in order to maintain the consistency of the estimates. The mean and the variance of the new Gaussian distribution represent the new estimate and the confidence (or certainty) associated with this new estimate respectively.

### E. Optimzation of Communication Overhead

Optimization of communication overhead is of prime importance in resource constrained and bandwidth-limited WSNs. The block *Decision Making* in Fig.1 is involved in this optimization mechanism of the proposed scheme. This module makes a trade-off between energy requirement and accuracy of the aggregated results.

To reduce the communication overhead, each node in the network communicates its computed estimate only when the estimate can bring a significant change in the estimates of its neighbors. For this purpose, each node stores the most recent value of the estimate it has received from each of its neighbors in a table. Every time a node computes its new estimate, it checks the difference between its newly computed estimate with the estimates of each of its neighbors. If this difference exceeds a pre-set threshold for any of its neighbors, the node broadcasts its newly computed estimate. The determination of this threshold is crucial as it has a direct impact on the level of accuracy in the global estimate and the energy expenditure in the WSN. A higher overhead due to message broadcast is optimized by maintaining two-hop neighborhood information in each node in the network [3]. This eliminates communication of redundant messages. This is illustrated in the following example.

Suppose that nodes *A*, *B* and *C* are in the neighborhood of each other in a WSN. Let us assume that node *A* makes a local measurement and this changes its global estimate. After combining this estimate with the other estimates of its neighbors as maintained in its local table, node *A* decides to broadcast its new estimate. As node *A* broadcasts its computed global estimate, it is received by both nodes *B* and *C*. If this broadcast estimate changes the global estimate of node *B* too, then it will further broadcast it to node *C*, as node *B* is unaware that the broadcast has changed the global estimate of node *C* also. Thus the same information is propagated in the same set of nodes in the network leading to a high communication overhead in the network.

To avoid this message overhead, every node in the network maintains its two-hop neighborhood information. When a node receives information from another node, it not only checks the estimate values of its immediate neighbors as maintained in its table but also it does the same for its two-hop neighbors. Thus in the above example, when node *B* receives information from node *A*, it does not broadcast as it understands that node *C* has also received the same information from node *A*, since node *C* is also a neighbor of node *A*. The two-hop neighborhood information can be collected and maintained by using algorithms as proposed in [11].

The choice of the threshold value is vital to arrive at an effective trade-off between the energy consumed for computation and the accuracy of the result of aggregation. For a proper estimation of the threshold value, some idea about the degree of dynamism of the physical process being monitored is required. A more dynamic physical process puts a greater load on the estimation algorithm thereby demanding more energy for the same level of accuracy [3]. If the user has no information about the physical process, he can determine the level of accuracy of the aggregation and the amount of energy spent dynamically as the process executes.

*F. Security of the Protocol*

The security module of the proposed scheme assumes that the sensing results for a set of sensors in the same neighborhood follows a normal (Gaussian) distribution. Thus, if a node receives estimates from one (or more) of its neighbors that deviates from its own local estimate by more than three times its standard deviation, then the neighbor node is suspected to have been compromised or failed. In such a scenario, the node that first detected such an anomaly sends a broadcast message to each of its neighbors requesting for the values of their estimates. If the sensing result of the suspected node deviates significantly (i.e., by more than three times the standard deviation) from the observation of the majority of the neighbor nodes, then the suspected node is detected as malicious. Once a node is identified as malicious, a broadcast message is sent in the neighborhood of the node that detected the malicious node and the suspected node is isolated from the network activities.

However, if the observation of the node does not deviate significantly from the observations made by the majority of its neighbors, the suspected node is assumed to be not malicious. In such a case, the estimate sent by the node is incorporated in the computation of the new estimate and a new global estimate is computed in the neighborhood of the node.

IV. SIMULATION RESULTS

In this section, we describe the simulations that have been performed on the proposed scheme. As the proposed algorithm is an extension of the algorithm presented in [3], we present here the results that are more relevant to our contribution, i.e., the performance of the security module. The results related to the energy consumption of nodes and aggregation accuracy for different threshold values (Section III E) are presented in detail in [3] and therefore these are not within the scope of this work.

In the simulated environment, the implemented application accomplishes temperature monitoring, based on network simulator (*ns-2*) and its sensor network extension *Mannasim* [12]. The nodes sense the temperature continuously and send the maximum sensed temperature only when it differs from the last data sent by more than 2%. In order to simulate the temperature behaviour of the environment, random numbers are generated following a Gaussian distribution, taking into consideration standard deviation of 1°C from an average temperature of 25°C. The simulation parameters are presented in Tab.1.

TABLE I. SIMULATION PARAMETERS

| Parameter | Value |
| --- | --- |
| No. of nodes | 160 |
| Simulation time | 200 |
| Coverage area | 120 m * 120 m |
| Initial energy in each node | 5 Joules |
| MAC protocol | IEEE 802.11 |
| Routing algorithm | None |
| Node distribution | Uniform random |
| Transmission power of each node | 12 mW |
| Transmission range | 15 m |
| Node capacity | 5 buffers |
| Energy spent in transmission | 0.75 W |
| Energy spent in reception | 0.25 W |
| Energy spent in sensing | 10 mW |
| Sampling period | 0.5 s |
| Node mobility | Stationary |

To evaluate the performance of the security module of the proposed algorithm, two different scenarios are simulated. In the first case, the aggregation algorithm is executed in the nodes without invoking the security module to estimate the energy consumption of the aggregation algorithm. In the second case, the security module is invoked in the nodes and some of the nodes in the network are intentionally compromised. This experiment allows us to estimate the overhead associated with the security module of the algorithm and its detection effectiveness.

It is observed that delivery ratio (ratio of the packets sent to the packets received by the nodes) is not affected by invocation of the security module. This is expected, as the packets are transmitted in the same wireless environment, introduction of the security module should not have any influence on the delivery ratio.

Regarding energy consumption, it is observed that the introduction of the security module has introduced an average increase of 105.4% energy consumption in the nodes in the network. This increase is observed when 20% of the nodes chosen randomly are compromised intentionally when the aggregation algorithm was executing. This increase in energy consumption is due to additional transmission and reception of messages after the security module is invoked.

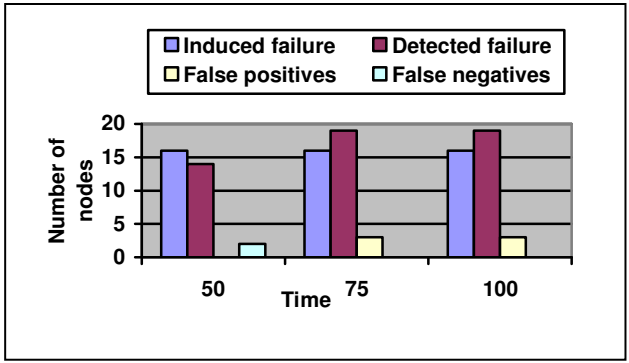

Figure 2. Detection effectiveness with 10% nodes in the network faulty

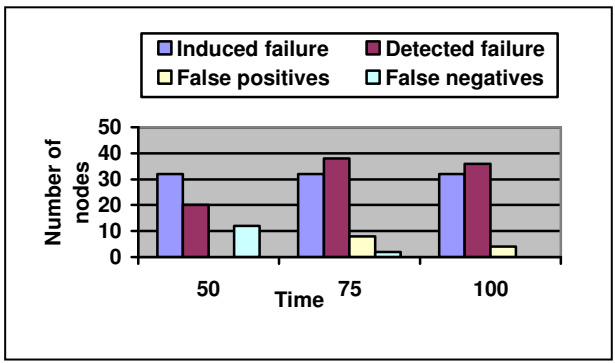

Figure 3. Detection effectiveness with 20% nodes in the network faulty

To evaluate the detection effectiveness of the security scheme, further experiments are conducted. For this purpose, different percentage of nodes in the network is compromised and the detection effectiveness of the security scheme is evaluated. Fig.2 and Fig.3 present the results for 10% and 20% compromised node in the network respectively. In these diagrams, the false positives refer to the cases where the security scheme wrongly identifies a sensor node as faulty while it is actually not so. False negatives, on the other hand, are the cases where the detection scheme fails to identify a sensor node which is actually faulty. It is observed that even when there are 20% compromised nodes in the network the scheme has a very high detection rate with very low false positive and false negative rate. The results show that the proposed mechanism is quite effective in detection of failed and compromised nodes in the network.

V. CONCLUSIONS

Building and deploying WSNs, especially in environments where there will be large number of nodes is a particularly complex task. Aggregation applications in these dense networks are extremely challenging tasks considering the computational and bandwidth constraints in these networks. In this paper, we have proposed an energy-efficient aggregation algorithm for WSNs that is secure and robust against malicious insider attack launched by compromised or faulty node(s). Simulations carried out on the proposed algorithm have demonstrated the effectiveness of the security module of the scheme.